\documentclass[a4paper,fleqn,useAMS,usenatbib]{mnras}

\pdfoutput=1

\usepackage{mathptmx}

\usepackage[T1]{fontenc}
\usepackage{ae,aecompl}

\usepackage[dvips]{graphicx}
\usepackage{amsmath}
\usepackage{amsfonts}
\usepackage{amssymb}
\usepackage{comment}
\usepackage[dvipsnames]{xcolor}
\usepackage[%
        ]{hyperref}
        
\hypersetup{
	colorlinks=true,	
	urlcolor=MidnightBlue,		
	pdfpagelabels=true,
	hypertexnames=true,
	plainpages=false,
	naturalnames=true,
	pdftitle={The Exomoon Corridor},    
	pdfauthor={Kipping},     
	linkcolor=WildStrawberry,          
	citecolor=ForestGreen,        
}

\newcommand{\kepler}{{\it Kepler}}

\newcommand{\pdf}{\mathrm{Pr}}

\title[The Exomoon Corridor]{
The Exomoon Corridor: Half of all exomoons exhibit TTV frequencies within a narrow window due to aliasing
}
\author[David Kipping]{David Kipping$^{1,2}$\thanks{E-mail:
\href{mailto:dkipping@astro.columbia.edu}{dkipping@astro.columbia.edu}}\\
$^{1}$Dept. of Astronomy, Columbia University, 550 W 120th Street, New York NY 10027\\
$^{2}$Center for Computational Astrophysics, Flatiron Institute, 162 5th Av., New York, NY 10010}

\date{Accepted . Received ; in original form }

\pubyear{2020}

\begin{document}
\label{firstpage}
\pagerange{\pageref{firstpage}--\pageref{lastpage}}
\maketitle

\begin{abstract}
Exomoons are expected to produce potentially detectable transit timing
variations (TTVs) upon their parent planet. Unfortunately, distinguishing
moon-induced TTVs from other sources, in particular planet-planet
interactions, has severely impeded its usefulness as a tool for identifying
exomoon candidates. A key feature of exomoon TTVs is that they will always
be undersampled, due to the simple fact that we can only observe the TTVs once
per transit/planetary period. We show that it is possible to analytically
express the aliased TTV periodicity as a function of planet and moon period.
Further, we show that inverting an aliased TTV period back to a true moon
period is fraught with hundreds of harmonic modes. However, a unique aspect
of these TTV aliases is that they are predicted to occur at consistently
short periods, irrespective of what model one assumes for the underlying
moon population. Specifically, 50\% of all exomoons are expected to induce
TTVs with a period between 2 to 4 cycles, a range that planet-planet
interactions rarely manifest at. This provides an exciting and
simple tool for quickly identifying exomoons candidates and brings the
TTV method back to the fore as an exomoon hunting strategy. Applying this
method to the candidate, Kepler-1625b i, reveals that its TTV
periodicity centers around the median period expected for exomoons.
\end{abstract}


\section{Introduction}

One of the first methods proposed to look for moons of extrasolar planets,
exomoons, is through transit timing variations (TTVs;
\citealt{sartoretti:1999}). The amplitude of these variations is proportional
to satellite-to-planet mass ratio, $(M_S/M_P)$, multiplied by the moon's
semi-major axis ($a_S$), and can anywhere from less than a second to up to an
hour \citep{impossible}.

Unfortunately, the detection of such signals is frustrated by two major
obstacles. First, planet-planet interactions are a frequent source of TTVs
(e.g. \citealt{holman:2010,nesvorny:2014,hadden:2017}), making it difficult to
assess whether a given TTV signal is truly caused by a satellite (e.g. see
\citealt{szabo:2013,HEK4}). Second, any satellite orbiting within the Hill sphere will
have an orbital period much shorter than the planetary period - which
represents the sampling rate via transits. Accordingly, the signal is
undersampled, as first proved in \citet{kipping:2009a}. A key conclusion
of that work was that ``the period of the exomoon cannot be reliably determined
from TTVs, only a set of harmonic frequencies''.

Despite the awareness of this issue since 2009, there has been no effort
(that we are aware of) to actually predict what this aliasing looks like. This
is important because characteristics of this aliasing may present a means of
distinguishing, even if only in a probabilistic sense, exomoon-induced TTVs versus
planet-planet interactions. At the most basic level, such an investigation
provides at lease some insight as to how exomoons should be expected to manifest
in the frequency domain of TTVs.

This is perhaps a product of the availability of other proposed methods to
distinguish between moons and planets in the literature at the time. For
example, several authors had expected that transits of the moons themselves
would be likely detected in conjunction with the TTVs \citep{brown:2001,
szabo:2006,simon:2007}, which would indeed permit a unique inversion
\citep{luna:2011}, as well offering the exomoon radius. Additionally, transit
\textit{duration} variations (TDVs) had been proposed as an additional
observable which could uniquely identify exomoons \citep{kipping:2009a,
kipping:2009b} and resolve the exomoon period. Thus, there was some optimism
that moons with Mars-to-Earth like masses should be detectable with \kepler\
and other facilities \citep{simon:2009,kepler:2009,awiphan:2013,simon:2015}.

Unfortunately, the stark reality is that these methods have not succeeded in
discovering a catalog of exomoons over the past decade.
This is likely not because some of fundamental flaw in the proposed techniques,
but rather a product of the fact Mars-to-Earth mass moons are apparently quite
rare around the known transiting planets \citep{HEK5,HEK6}. Moons akin to
the Galilean satellites, with mass ratios of $10^{-4}$, are simply far too
small to be detectable through TDVs with \kepler\ \citep{kipping:2009a,awiphan:2013}.
Similarly, their transits would be almost always undetectable due to \kepler's
poor completeness for long-period sub-Earths \citep{christiansen:2016}.
However, widely-separated low-mass moons can still produce plausibly detectable
TTVs, which can be seen by re-parameterizing the moon TTV equation and employing
canonical units:

\begin{align}
\mathrm{TTV}_{\mathrm{amp.}} &= (34.2\,\mathrm{sec})\,f
\Bigg( \frac{M_P/M_{\star}}{M_J/M_{\odot}} \Bigg)^{1/3}
\Bigg( \frac{M_S/M_P}{10^{-4}} \Bigg)
\Bigg( \frac{P_P}{\mathrm{years}} \Bigg),
\end{align}

where $f$ is the semi-major axis of the satellites, $a_S$, divided by the planet's
Hill radius, $R_H$. On this basis, intermediate-mass moons (sub-Earth but larger
than the Galilean moons) could have evaded efforts focussed on the photometric detection
of exomoon transits (e.g. \citealt{HEK5}), yet be presenting
detectable-but-ambiguous TTV signatures in the current data.

In what follows, we therefore revisit the theory underpinning exomoon TTVs, with
particular attention to the aliasing effect thus far ignored.

\section{Theory}

\subsection{Why moons are always undersampled}

Consider a satellite orbiting a planet with a semi-major axis, $a_S$, equal to
some fraction, $f$, of the planet's Hill radius, $R_H$. From
\citet{kipping:2009a}, one can show that the exomoon period is related to the
planet's period via the relationship

\begin{align}
P_S \simeq P_P \sqrt{\frac{f^3}{3}},
\label{eqn:stable}
\end{align}

under the assumption that $M_P \ll M_{\star}$ and $M_S \ll M_P$. As a frequency,
this becomes $\nu_S = \sqrt{3/f^3}/P_P$. Since we only
measure a TTV once per orbital period of the planet, then the sampling rate
is $\nu_P = 1/P_P$. This in turn means that the Nyquist frequency is
$\nu_{\mathrm{Nyquist}} \equiv 0.5\nu_P = 0.5/P_P$, above
which any frequency will be undersampled and become aliased. It is easy to see
that this is indeed always the case for a stable bound satellite, since

\begin{align}
\sqrt{3/f^3}/P_P > 0.5/P_P
\end{align}

for all $f<2^{2/3}3^{1/3}=2.289$. Given that moons must lie within the
Hill sphere for stability requirements (i.e. $f<1$), then this establishes that
exomoons will always manifest as an undersampled TTV signature.

\subsection{Calculating the aliased frequency}

Given that the moon signal will be undersampled, it will still manifest in the
TTVs but as an aliased signal i.e. a frequency not equal to the true frequency
of the underlying signal. The sampling frequency, $\nu_P$, will mix with the
satellite frequency, $\nu_S$, to produce an aliased frequency we denote
as $\nu_{\mathrm{TTV}}$ (which will be below the Nyquist rate).

In searching for a solution to this problem, we are guided by the analogy that
can be drawn between this problem and one described in a seminal paper by
\citet{dawson:2010}. In that paper, the authors analyzed the radial velocity
(RV) measurements of the several stars, including 55-Cancri, and argued that an
aliasing effect was misleading previous analyses of the inferred periodicities.
In particular, radial velocity surveys at the time commonly observed each star
no more than once per night, leading to a sampling rate of approximately one
day. Thus, planets with periods less than 2\,days would be undersampled and
manifest as aliases in the frequency analysis of such RVs. \citet{dawson:2010}
argued that this was likely the case for planet 55-Cancri e, a signal with an
apparent period of 2.8\,days argued to be an alias of a true period at 0.74\,d.
This was later confirmed through transit observations of the system using
\textit{Spizter} by \citet{demory:2011}.

We follow \citet{dawson:2010} and \citet{mcclellan:1999} in considering the
effect of undersampling on our closely related problem. \citet{mcclellan:1999} showed
that an undersampled signal (in our case the TTV signal) is exactly fit by
frequencies satisfying

\begin{align}
\nu &= \nu_S \pm m \nu_P = \frac{1}{P_S} \pm m \frac{1}{P_P},
\label{eqn:aliases}
\end{align}

where $m$ is a positive real integer. If we compute a periodogram/power
spectra of the TTV signal, any aliased negative frequencies in the above
will still manifest, such that our observed aliased peaks would occur at

\begin{align}
\nu &= \Big|\frac{1}{P_S} \pm m \frac{1}{P_P}\Big|.
\end{align}

In order for these frequencies to be included in the periodogram, they
must satisfy $\nu < \nu_{\mathrm{Nyquist}}$ (since periodograms
are conventionally not computed for frequencies above the Nyquist rate),
and thus we require

\begin{align}
\Big|\frac{1}{P_S} \pm m \frac{1}{P_P}\Big| < \frac{0.5}{P_P}.
\end{align}

We are free to work in whatever units we wish, so multiplying through $P_P$,
the above becomes

\begin{align}
\Big|\frac{P_P}{P_S} \pm m\Big| < \frac{1}{2}.
\end{align}

Since $P_P > P_S$ (see Equation~\ref{eqn:stable}), then the majority of choices
of $m$ will not satisfy the equation above. Let's say that the ratio was 13.37,
as an arbitrary example (this is the ratio for the Earth-Moon). Only the choice
$m=13$ with a negative sign will satisfy the expression, as $m=12$ would lead
to 1.37 inside the modulus and $m=14$ would lead to 0.63. In fact, since
$P_P > P_S$ for all stable moons, then there will always only be one valid
solution to the above which occurs when $\pm$ takes the negative sign and we
have

\begin{align}
m &= \mathrm{round}[P_P/P_S].
\end{align}

Ergo, the aliased TTV frequency will always appear uniquely at

\begin{align}
\nu_{\mathrm{TTV}} &= \nu_S - \mathrm{round}\big[\tfrac{\nu_S}{\nu_P}\big] \nu_P,
\end{align}

giving an observed TTV periodicity of

\begin{align}
P_{\mathrm{TTV}} &= \frac{1}{ \tfrac{1}{P_S} - \mathrm{round}\big[\tfrac{P_P}{P_S}\big] \tfrac{1}{P_P} }.
\label{eqn:Pobs}
\end{align}

\begin{figure*}
\begin{center}
\includegraphics[width=18.0cm,angle=0,clip=true]{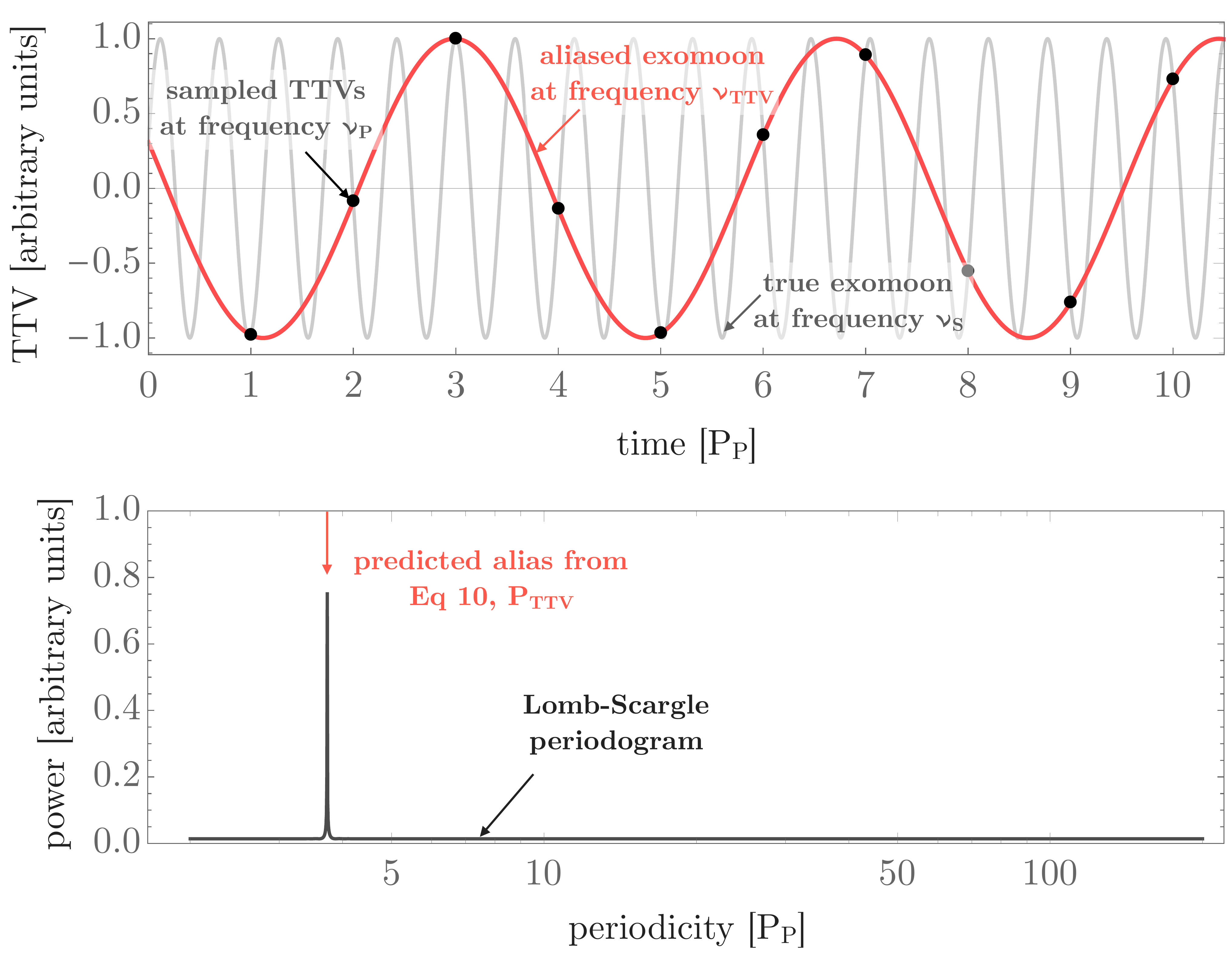}
\caption{
Example of TTV aliasing for a Hill radius exomoon. The moon's TTV
signal (shown in gray) has frequency above the Nyquist rate (this is
always true for moons) and thus is perfectly fit by the longer period
alias shown in red. The periodogram computed below shows the same
effect in the frequency domain.
}
\label{fig:example}
\end{center}
\end{figure*}

As an illustrative example of the aliasing effect, Figure~\ref{fig:example}
shows an example of 10 epochs of a planet with a Hill radius exomoon. The
true waveform, shown in gray, is only sampled once per planetary period
(black points) and is perfectly fit by the longer period alias shown in
red. This can be also be seen in periodogram of Figure~\ref{fig:example}.

\subsection{Going from observed period to exomoon period}

If one observes a TTV signal with a periodicity of $P_{\mathrm{TTV}}$, one now
needs to invert Equation~(\ref{eqn:Pobs}) back to the true period, $P_S$. We find
that any $P_S$ which satisfies the below expression will be an exact solution
to Equation~(\ref{eqn:Pobs}):

\begin{align}
P_S &= \Big(\frac{1}{P_{\mathrm{TTV}}} + n \frac{1}{P_P}\Big)^{-1},
\end{align}

where $n$ is a positive real integer. If we require that $f<1$ (the satellite
is inside the Hill sphere), then we can impose

\begin{align}
n > \frac{\sqrt{3}}{f^{3/2}} - \frac{P_P}{P_{\mathrm{TTV}}}.
\end{align}

For $f<1$ and $P_{\mathrm{TTV}} > 2P_P$ (the Nyquist periodicity), this
always excludes $n=1$ but leaves anything greater than or equal to 2 as
a possibility.

A maximum limit on $n$ could be proposed if one adopts some lower limit for
$P_S$. Io is perhaps the most extreme example of a large moon in close
proximity to its planet, residing within at 0.8\% of Jupiter's Hill radius. If
we take $f=0.01$ as a fiducial lower limit, then we find that one can fit
1731 unique solutions for $P_S$ in-between this limit and the Hill radius. This
highlights how challenging and multimodal the exomoon parameter space truly is.

\subsection{Probability distribution of TTV periods due to aliased moons}

We have now established that i) exomoons always induce undersampled TTVs which
manifest as an aliased TTV, and ii) inverting an aliased periodicity back to
the true exomoon period is highly multimodal. The last task we address here
considers what the probability distribution of these aliased TTV periods are
expected to be.

In order to address this, we first need to choose a probability distribution 
for the exomoon periods. It is convenient to work in units of planetary period
for the TTV periods and frequencies, since then the distribution of the
$P_P$ does not actually affect our results. However, we do still need to choose
some distribution for the moons themselves. To this end, we considered four
possible toy models:

\begin{itemize}
\item[{\textbf{A]}}]
Exomoon periods distributed such that they are uniform in frequency
space (i.e. uniform in $1/P_S$).
\item[{\textbf{B]}}]
Log-uniformly distribution for $f$, which is equivalent to a log-uniform
distribution for semi-major axis $a_S$ (and also $P_S$).
\item[{\textbf{C]}}]
Uniformly distribution for $f$, which is equivalent to a log-uniform
distribution for semi-major axis $a_S$ (but not in $P_S$).
\item[{\textbf{D]}}]
A linearly weighted distribution in moon semi-major axis such that
$\mathrm{Pr}(a_S) \propto a_S$.
\end{itemize}

Whilst the first three are broadly designed to be diffuse, uninformative
distributions, the last one encodes a selection bias effect that we are
more likely to detect wider orbit moons, since the TTV amplitude is itself
proportional to $a_S$ \citep{sartoretti:1999}.

\begin{figure*}
\begin{center}
\includegraphics[width=18.0cm,angle=0,clip=true]{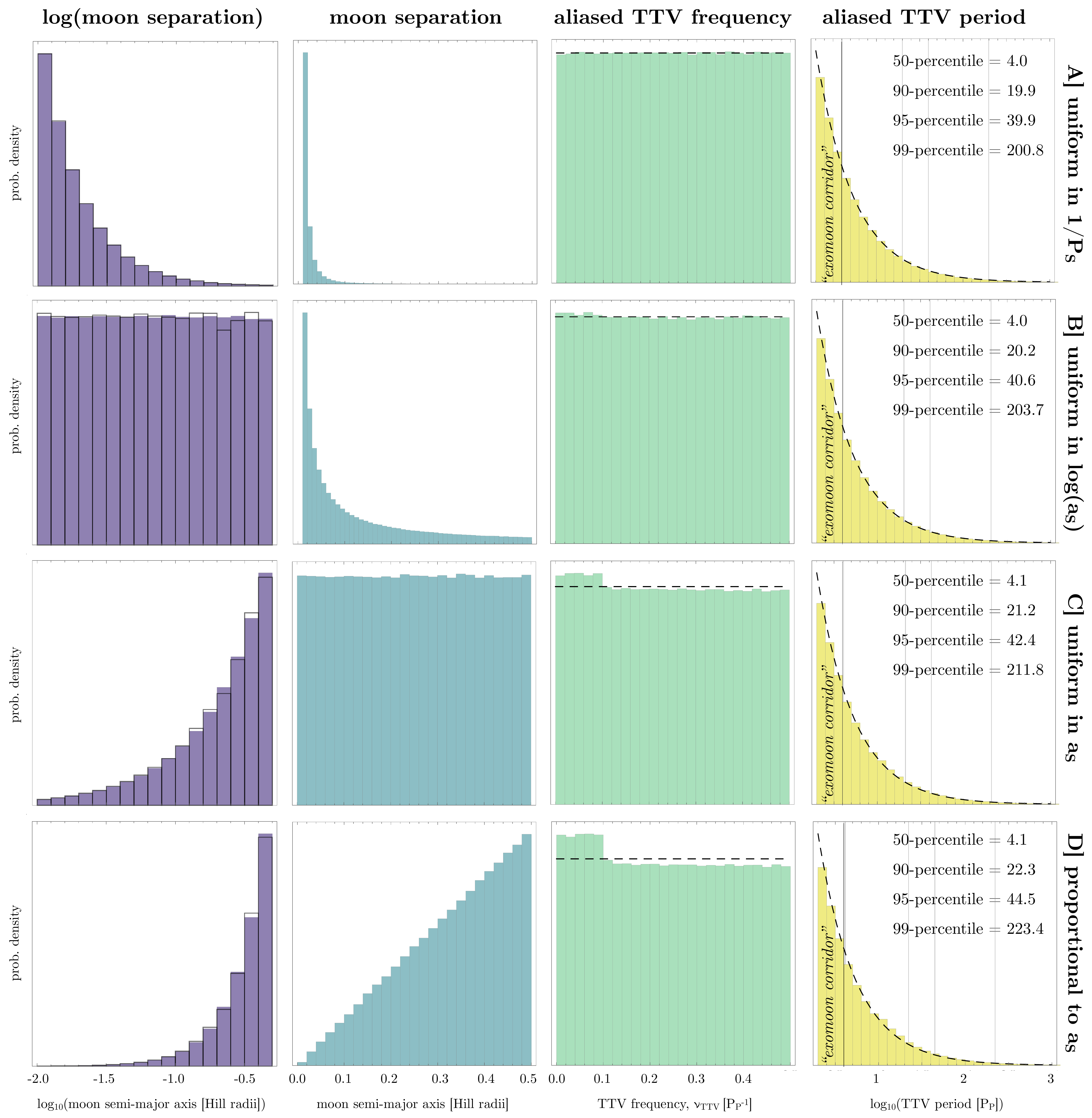}
\caption{
The distribution of four relevant exomoon TTV parameters (the four
columns) for four different assumed distributions of exomoons
(the four rows). Although the models produce wildly different
distributions for the moon separations (left columns), the
knock-on effect to the distribution of aliased TTVs is fairly
self-similar and thus robust.  Vertical grid lines show the median,
90\%, 95\% and 99\% quantiles. Black histograms overlaid on
the left column shows the same distribution but applying the filter
that the moons must manifest in the labelled ``exomoon corridor''.
Black dashed lines show an closed-form approximation given by
Equation~(\ref{eqn:formula}).
}
\label{fig:grid}
\end{center}
\end{figure*}

We assume a minimum allowed $f=0.01$ and a maximum of $f=0.5$
\citep{domingos:2006} and generated $10^6$ samples from these distributions.
The resulting distributions for $\log_{10} f$, $f$, $\nu_{\mathrm{TTV}}$ and
$\log_{10} P_{\mathrm{TTV}}$, for all four models, are shown in
Figure~\ref{fig:grid}.

As evident from the figure, there are very pronounced differences
between how the moons are distributed spatially amongst the four models.
Yet, despite this, the resulting distribution of aliased TTV frequencies
and periodicities are remarkably self consistent. In every experiment,
we recover an exponential-like distribution in $\log_{10} P_{\mathrm{TTV}}$
piling up at the Nyquist period. One can see from the quantiles
listed on the far right column that the median aliased TTV period is about
4 cycles. In other words, 50\% of all exomoons should be expected to
produce a TTV aliased period between 2 and 4 cycles, a region we
refer to as the ``exomoon corridor''.

From Figure~\ref{fig:grid}, we note that the TTV frequency panels exhibit
an approximately uniform distribution. If we adopt this as a formal
approximation, we can use it to transform into period-space and express a
closed-form expression for the period distribution:

\begin{equation}
\pdf(P)\,\mathrm{d}P =
\begin{cases}
2 P^{-2}\,\mathrm{d}P & \text{if } P>2 ,\\
0 & \text{otherwise } .
\end{cases}
\label{eqn:formula}
\end{equation}

This distribution is overplotted in the right-hand panels of
Figure~\ref{fig:grid}, where one can see that it provides good agreement to the
numerical experiments. The cumulative distribution function of
Equation~(\ref{eqn:formula}) can be trivially shown to yield a median of 4 - thus
recovering the observation that 50\% of the samples produce a period in the
range of 2 to 4 cycles.

\subsection{The exomoon corridor as a detection strategy}

Given that half of all moons manifest in the corridor, broadly
irrespective of their true underlying distribution, this means that any
TTV signal in this region could be potentially rapidly identified
as an exomoon candidate. If this strategy were implemented, it is
natural to ask whether the recovered moons would be in any way biased
towards a certain region of parameter space.

To investigate this, we took our four fake moon populations, and filtered
on only those samples which ended up in the corridor when considering
$P_{\mathrm{TTV}}$. This sub-sample, which represents about the half the
population, was then compared to the original injected population to see if
any systematic changes had occurred. This is visualized in Figure~\ref{fig:grid}
in the left column, where the colored histograms show the original, injected
population and the black outline histograms show the filtered subset. As
evident from the figure, the distributions are nearly identical and thus
selecting exomoons that live in the corridor will not noticeably bias the
recovery of the true population.

Although we have shown that half of all moons will appear in the corridor,
this would not be of much practical value if false-positives also preferentially
pile-up in this region. As discussed in the introduction of this paper,
the most common source of large TTVs is planet-planet interactions and thus
it is natural to ask what their $P_{\mathrm{TTV}}$ distribution looks like.
One can analytically predict $P_{\mathrm{TTV}}$ for planet-planet interactions,
which are typically dominated by the so-called ``super-period'' of circulating
line of conjunctions \citep{lithwick:2012} and to a lesser degree the
``chopping'' effect caused by the conjunctions themselves \citep{nesvorny:chop,
deck:2015}. However, this calculation is highly sensitive to precise period
ratios between planets, which cannot be assumed to be drawn from independent
distributions due to the mutual interactions themselves.

\begin{figure}
\begin{center}
\includegraphics[width=8.4cm,angle=0,clip=true]{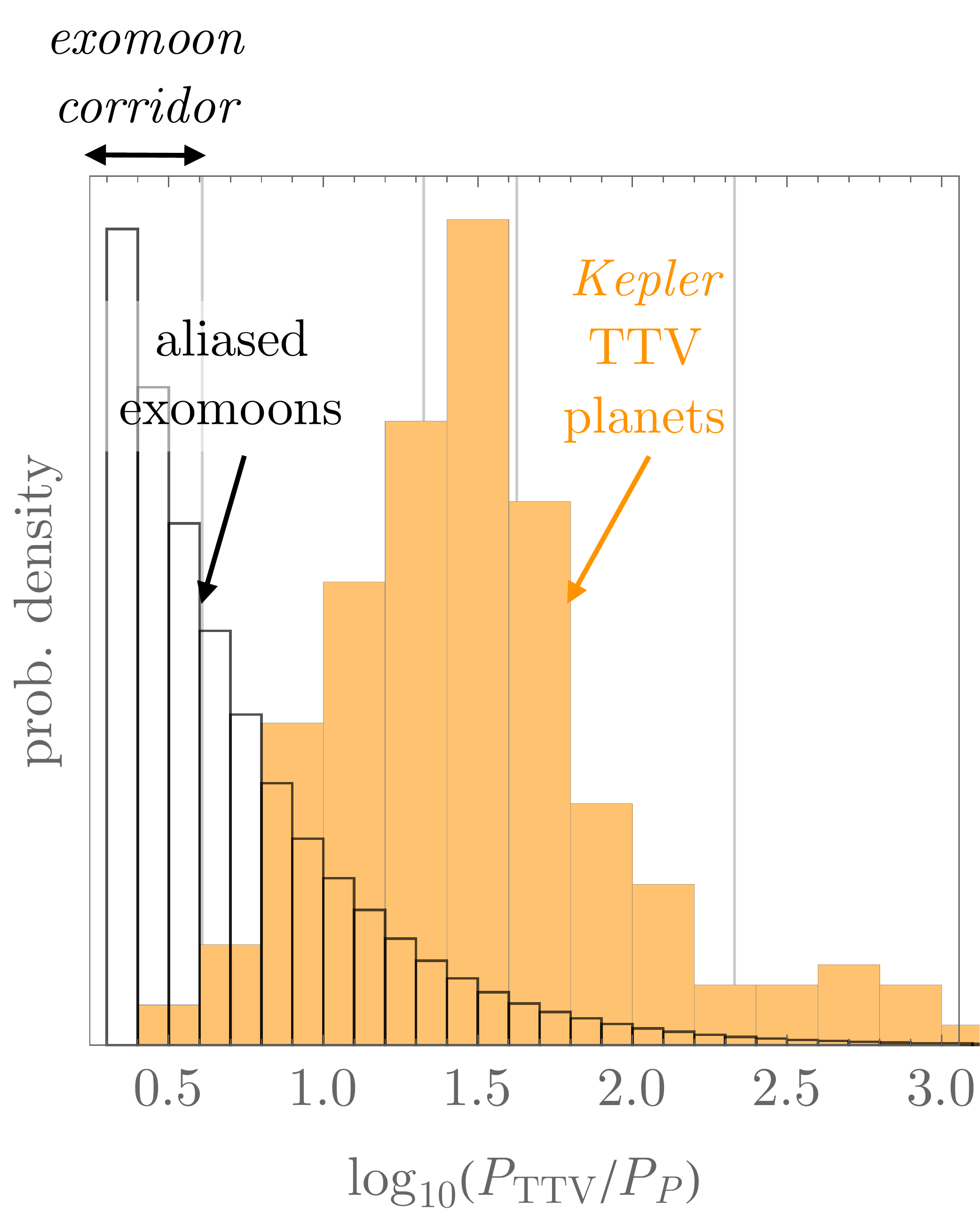}
\caption{
Comparison of the TTV periods for \kepler\ planet-planet pairs (orange)
exhibiting gravitational interactions \citep{hadden:2017}, versus the
predicted distribution of TTV periods caused by aliased exomoons (black).
Only ${\simeq}1$\% of TTV planets manifest in the ``exomoon corridor'' of
2-4 cycles.
}
\label{fig:kepler}
\end{center}
\end{figure}

Instead, we took the largest catalog of confirmed TTV planets available in the
literature from \citet{hadden:2017}. Specifically, their Table~2 lists 90
adjacent planet pairs with mutual interactions, from which we calculate the
dominant super-period as $P_{\mathrm{TTV}}$ for 180 planets. A histogram of
these periods is shown in Figure~\ref{fig:kepler}, where one can clearly see a
quite distinct distribution from those shown earlier in Figure~\ref{fig:grid}.

Only two out of 180 TTV planets exhibit a TTV super-period in the exomoon
corridor i.e ${\simeq}1$\%. Given that 50\% of moons manifest in the corridor,
it might be tempting to assign a ${\sim}$50:1 odds ratio that any signal in
this region is moon-like. However, such a simple analysis ignores the
observational bias that planet-planet interactions are generally much larger
than that due to moons, sometimes of order of many days \citep{koi142}, and
thus would be overrepresented versus this naive fraction. An accurate
false-positive rate would require knowledge of the mass distribution of both
interacting planets and exomoons, which for the latter we certainly do not
have. Further, there are expected to be other sources of false-positives
besides planet-planet interactions, such as spurious TTVs induced by stellar
activity \citep{ioannidis:2016}, and transit-phase sampling effects
\citep{kipping:2011,szabo:2013}.

Nevertheless, it is encouraging that planet-planet interactions are expected to
rarely produce TTVs in the exomoon corridor. Signals in the region are thus
immediately interesting for exomoon follow-up analysis, such as detailed
photodynamical modeling \citep{luna:2011}. Since such an analysis is
computationally expensive (only $\sim60$ out of the several thousand \kepler\
candidates have been analyzed thus far; \citealt{HEK5}), the exomoon corridor
provides a useful means of rapidly prioritizing interesting candidates for
subsequent investigation.

\subsection{Application to Kepler-1625b}

As a demonstration of the described technique on an exomoon candidate, we
here apply the outlined method to Kepler-1625b i. However, using the 4 available
transit times reported in \citep{teachey:2018}, a naive periodogram analysis
is problematic. This is because the problem is underdetermined with 5 unknown
parameters; 2 linear ephemeris parameters ($P$ and $\tau$) + 3 sinusoid
parameters ($A_{\mathrm{TTV}}$, $P_{\mathrm{TTV}}$ and $\phi_{\mathrm{TTV}}$).

A common approach would be to fit a linear ephemeris first, then compute the
four residuals to that fit - which defines TTVs. At this point, the problem
might seem now overdetermined, since we'd next model the 4 TTVs with a
3-parameter sinusoid. However, this approach is formally wrong since inference
of the sinusoid terms would be strictly conditional upon the previously adopted
ephemeris. Critically, that ephemeris has uncertainty, it is not known to
infinite precision - especially with just 4 available transits. The
inter-parameter degeneracies, described by the mutual covariances, mean that
this simple approach will not in general yield accurate solutions.

Progress can be made in underdetermined problems through regularization, or
in Bayesian parlance, the use of priors. As an example, in machine learning,
Lasso regularization is a common strategy to further constrain a problem
by appending a new term to the merit function that penalizes large parameter
values \citep{lasso:1995}. Priors have a similar effect to regularization,
since they add too extra constraints/information into the inference and in
many ways can be thought of as an equivalent approach.

\begin{figure}
\begin{center}
\includegraphics[width=8.4cm,angle=0,clip=true]{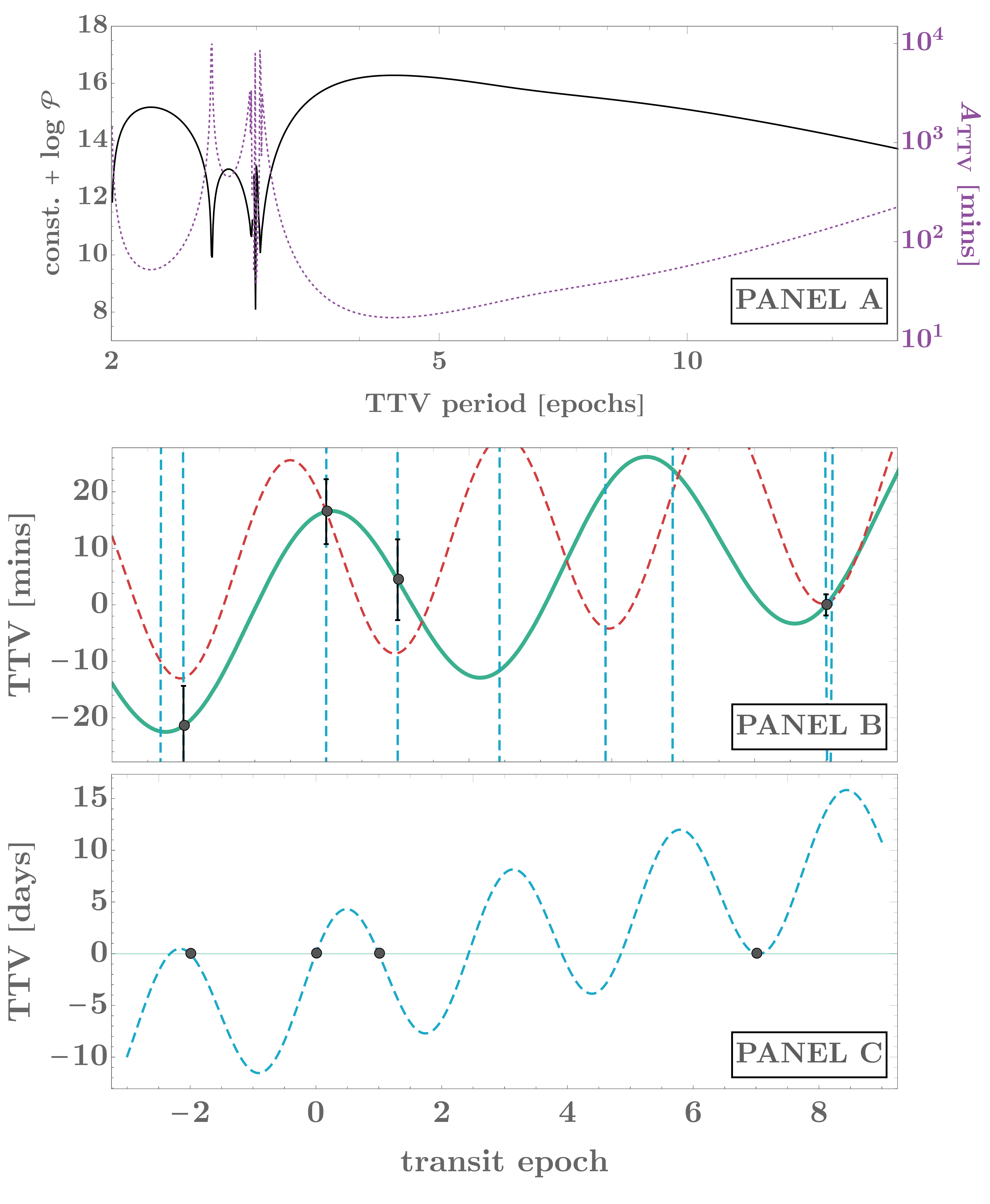}
\caption{
Panel A: Periodogram of the TTVs, with a broad peak occurring at 4.4 cycles
(corresponding to $\Delta\chi^2=17.7$) - approximately
the median TTV period that would be expected from an exomoon.
Panel B: Observed transit times of Kepler-1625b from \citet{teachey:2018}
after subtracting the maximum a-posteriori ephemeris, to yield TTVs.
The maximum a-posteriori sinusoid is shown in green with a period of 4.4 cycles.
The red line shows the one ``bad'' period for which a perfect fit could not
be obtained.
Panel C: A zoom-out of the TTVs showing an example of a perfect fit with
a ridiculous amplitude (blue line, which is also present in Panel B). Such
fits are strongly disfavoured in our final periodogram through the use of regularization.
}
\label{fig:1625}
\end{center}
\end{figure}

To gain some insight into the appropriate regularization here, we first
attempted a naive periodogram despite the overdetermined nature (see
Figure~\ref{fig:1625}A). Using a log-uniform grid of candidate TTV periods, we
fit the now four-parameter model (since $P_{\mathrm{TTV}}$) at each period -
yielding in almost every case a perfect fit in a $\chi^2$-sense. The only
exception to this was at $P_{\mathrm{TTV}}=3$ cycles, which curiously was the
only period unable to obtain a perfect fit, with $\chi^2=4.75$ (red line in
Figure~\ref{fig:1625}B). On the face of it, this periodogram teaches us very
little - every period is as good as any other (with one exception). However,
if we plot the amplitude of the maximum likelihood sinusoids at each period
(see Figure~\ref{fig:1625}A purple line), one sees extremely large amplitudes
often being invoked to explain the data. For example, at
$P_{\mathrm{TTV}}=2.65$\,cycles, the amplitude is nearly 7\,days (blue line
in Figure~\ref{fig:1625}B and C). Given that the RMS of the four data
points is 16\,minutes, it seems highly improbable that four randomly drawn
points from a 7\,day amplitude sinusoid would be so small. Much like Lasso
regularization then, we seek to instruct our inference to prefer low amplitude
TTVs over high amplitude ones. This can be formally encoded by adding a
log-uniform prior to the amplitude of the form $\mathrm{Pr}(A) \sim
1/A_{\mathrm{TTV}}$. We can then add the logarithm of this prior onto the
log-likelihood from a simple least squares normal) likelihood function to
express a log-posterior probability (unnormalized) for each period grid point.
A similar approach has been used for Bayesian radial velocity periodograms
(e.g. see \citealt{gregory:2007}).
The resulting periodogram is shown in Figure~\ref{fig:1625} in black.

The maximum a-posteriori peak is broad, which is not surprising given the poor
constraints, but centers on 4.4\,cycles. In the context of this paper, this
occurs very close to the median period expected from an exomoon. We also note
that the existence of TTVs is quite secure at $\Delta\chi^2=17.5$; an issue
which should not be conflated with the distinct issue of parameter
determination. These two points thus show that Kepler-1625b is indeed
consistent with the exomoon hypothesis claimed by \citet{teachey:2018}, using
just the transit times alone. To share this maximum a-posteriori best model,
we report $A_{\mathrm{TTV}} = 17.08$\,mins, $P_{\mathrm{TTV}} = 4.405$\,cycles,
$\phi_{\mathrm{TTV}} = 1.5478$\,rads, $P = 287.37221$\,days and
$\tau = 2456043.95957$\,BJD$_{\mathrm{UTC}}$, where we define the median
\kepler\ epoch as the reference ($0^{\mathrm{th}}$) epoch.

\section{Discussion}

We have shown that TTVs due exomoons will always manifest as longer period
aliases in the frequency analysis of such observations. It is possible
to analytically express the exact relationship between the exomoon period
and the aliased frequency, as given by Equation~(\ref{eqn:Pobs}) in this work.
Converting these aliased periods back to unique exomoon periods appears
intractable with TTVs alone, due to the many hundreds of harmonic solutions
which can fit the data. Thus, unique solutions to exomoon data will surely
depend on auxiliary information such as transits \citep{brown:2001,szabo:2006,
simon:2007,luna:2011} or TDVs \citep{kipping:2009a,kipping:2009b,awiphan:2013}.

However, we note that injecting a variety of radically different distributions
for the exomoon population into our formula leads to nearly identical resulting
distributions for the aliased TTV period (given by Equation~\ref{eqn:formula}).
Specifically, exomoons produce aliased TTVs that preferentially peak close to
the Nyquist periodicity of 2 cycles. Indeed, we estimate that 50\% of all
exomoons will manifest in a narrow TTV period corridor of 2 to 4 cycles, broadly
independent of the assumed exomoon distribution.

This remarkable feature provides an immediate and powerful tool for
identifying exomoon candidates. Any TTV signal with a strong peak below 4
cycles is an excellent target for further analysis. Naturally, this
information alone does not demonstrate the moon hypothesis and so we caution
readers to not abuse this tool. Although planet-planet interactions less
frequently manifest at such short periodicities, chopping signals (due to
conjunctions) can mimic shorter term variability \citep{nesvorny:chop,
deck:2015}, as well as false positives from spots for example 
\citep{ioannidis:2016}. A detailed investigation of the power spectra of these
effects is encouraged for future work. Similarly, planets with
very large TTVs would provide non-uniform sampling that may offer access
to higher frequencies \citep{murphy:2013,vanderplas:2018}, and thus would also
be a worthwhile topic of further investigation.

Nevertheless, this work provides a
simple and straight forward tool to quickly identifying high priority exomoon
candidates suitable for further analysis with tools such as photodynamical
modeling.

\section*{Acknowledgments}

DMK is supported by the Alfred P. Sloan Foundation.

\section*{Data Availablity}

The data underlying this article are available via Columbia Academic Commons, at https://doi.org/10.7916/D8795NHS. The datasets were derived from sources in the public domain \citep{teachey:2018}.


%

\bsp
\label{lastpage}
\end{document}